\documentstyle[12pt]{article}

\newcommand{\bd}{\begin{document}}
\newcommand{\ed}{\end{document}}
\newcommand{\bc}{\begin{center}}
\newcommand{\ec}{\end{center}}
\newcommand{\be}{\begin{eqnarray}}
\newcommand{\ee}{\end{eqnarray}}
\renewcommand{\thefootnote}{\alph{footnote}}
\newcommand{\se}{\section}
\newcommand{\sse}{\subsection}
\newcommand{\bi}{\bibitem}
\renewcommand{\thefootnote}{\fnsymbol{footnote}}
\def\figcap{\section*{Figure Captions\markboth
     {FIGURECAPTIONS}{FIGURECAPTIONS}}\list
     {Figure \arabic{enumi}:\hfill}{\settowidth\labelwidth{Figure 999:}
     \leftmargin\labelwidth
     \advance\leftmargin\labelsep\usecounter{enumi}}}
\let\endfigcap\endlist \relax
\input psfig

\begin{document}

\begin{titlepage}

 \vskip 0.5in
 \null
\begin{center}
 \vspace{.15in}
{\Large {\bf Analysis of \boldmath{$B_{s} \rightarrow KK$} decays
in the PQCD
}}\\
\vspace{1.0cm}  \par
 \vskip 2.1em

{\Large\bf Chuan-Hung Chen} %

\vspace{0.7cm}

   {\Large \sl Department of Physics, National Cheng Kung University}
\\   {\Large \sl  $\ $Tainan, Taiwan,  Republic of China }

 \par \vskip 5.3em

\date{\today}
 {\Large\bf Abstract}

\end{center}
We study $B_{s}\rightarrow KK$ decays in the framework of the PQCD.
 We show that the branching ratios of
 $B_{s}\rightarrow (K^+ K^-, K^0 \bar{K}^0)$ are
about $(22.43, 25.78) \times 10^{-6}$
 for $\phi_{3}(\gamma)\simeq72^0$, which are consistent with the
 model-independent estimations.
We find that the typical scale of the decays is near $1.7$ GeV. We
also point out that the induced strong phase $\delta$ is around
$207^0$ so that the direct CP asymmetry of $B_{s}\rightarrow
K^+K^-$ could reach $15\%$.\\[0.3cm]

\end{titlepage}


The study of charmless B decays has an enormous progress since many decays
such as those with exclusive pseudoscalar-pseudoscalar final states $%
(B\rightarrow PP)$ were measured at $e^{+}e^{-}$ machines by CLEO \cite{CLEO}%
, BABAR \cite{BCP41} and Belle \cite{BCP42}, respectively. From
the search of B decays, we not only could test the origin of CP
violation in standard model (SM), which is the consequence of the
Cabibbo-Kobayashi-Maskawa (CKM ) quark-mixing matrix \cite{CKM}
but also verify various QCD approaches for nonperturbative
problems in exclusive decays. Recently, the proposals of using the
mixing-induced and direct CP asymmetries in $B_{d}\rightarrow
\pi^{+}\pi ^{-}$ and $B_{s}\rightarrow K^{+}K^{-}$
\cite{Fleischer2} or the branching ratios (BRs) of $B_{s}\to
K^+K^-$ and $K^0\bar{K}^0$ decays \cite {CQ} are suggested to
determine the angle $\phi _{3}$ or $\gamma $. Clearly,
it is important to give a detailed analysis on the decays of $%
B_{s}\rightarrow KK$.

It is known that one of the main theoretical uncertainties for studying
exclusive hadron decays is from the calculations of matrix elements. Usually
it is performed in the perturbative QCD (PQCD) approach developed by Brodsky
and Lepage (BL) \cite{LB1}. In the BL formalism, the nonperturbative part is
included in the hadron wave functions and the transition amplitude is
factorized into the convolution of hadron wave functions and the hard
amplitude of the valence quarks. However, with the BL approach, it has been
pointed out that perturbative evaluation of the pion form factor suffers a
non-perturbative enhancement from the end-point region with a momentum
fraction $x\to 0$ \cite{IL}. If so, the hard amplitude is characterized by a
low scale, such that the expansion in terms of a large coupling constant $%
\alpha _{s}$ is not reliable. Furthermore, more serious end-point
(logarithmic) singularities are observed in the twist-2 (leading-twist)
contribution to the $B\to \pi $ transition form factors \cite{SHB,ASY}. The
singularities even become linear while including the twist-3
(next-to-leading twist) wave function \cite{BF}. Because of these
singularities, it was claimed that form factors are dominated by the soft
dynamics and not calculable in the PQCD \cite{KR}.

In order to take care of the end-point singularities, $k_{T}$, the
transverse momentum of the valence-quark \cite{LS}, and threshold
resummations \cite{S0,CT} have to be introduced. The inclusion of $k_{T}$
will bring in large double logarithms $\alpha _{s}\ln ^{2}(k_{T}/M_{B})$
through radiative corrections. Therefore, these large logarithms should be
resumed in order to improve the perturbative calculation. Due to the
resummation \cite{LS,CS,BS}, the arisen distribution of $k_{T}$ exhibits a
suppression in the region with $k_{T}\sim O(\bar{\Lambda})$ \cite{KLS} and
the average of $k_{T}^{2}$ is enlarged up to around $\langle
k_{T}^{2}\rangle \sim \bar{\Lambda}M_{B}$ for $M_{B}\sim 5$ GeV;
consequently, the off-shellness of internal particles keeps being $O(\bar{%
\Lambda}M_{B})$ even in the end-point region. Moreover, due to the radiative
corrections of the weak vertex \cite{Li1}, another type of double logarithms
$\alpha _{s}\ln ^{2}x$ actually exists while $x\rightarrow 0$; and
therefore, these large corrections should be also resumed, called threshold
resummation \cite{S0,CT}, for justifying the perturbative expansion so that
the result leads to a strong Sudakov suppression at $x\to 0$. Hence,
including $k_{T}$ and threshold resummations, the end-point singularities
can be dealt with self-consistent in the PQCD.

In the literature, the applications of the modified PQCD (MPQCD)
\cite{MPQCD} to the processes of $B\rightarrow PP$, such as
$B\rightarrow K\pi $ \cite{KLS1}, $ B\rightarrow \pi \pi $
\cite{pipi}, $B\rightarrow KK$ \cite{CL} and $B\rightarrow
K\eta^{(\prime)}$ \cite{KS}, as well as that of $B\rightarrow VP$
such as $B\rightarrow \phi \pi $ \cite{Melic}, $B\rightarrow \rho
(\omega) \pi $ \cite{LY} and $B\rightarrow \phi K$
\cite{Mishima,CKL} have been studied and found that they are
consistent with the experimental data or limits. For a review on
the PQCD approach, we summary the characters of the MPQCD briefly
as follows:

\begin{itemize}
\item  Due to the introduction of $k_{T}$ and threshold resummations for
smearing the singularities, the $B\rightarrow \pi ,K$ form factors are still
dominated in the perturbative region with $\alpha _{s}/\pi <0.2$ \cite{KLS1}.

\item  There involve three scales in the MPQCD: the $M_{W}$ scale of the
electroweak (EW) interaction, the typical scale $t$ which reflects the
specific dynamics of the heavy meson decays, and the factorization scale of $%
\Lambda \simeq M_{B}-M_{b}$ with $M_{B}$ and $M_{b}$ being the B-meson and
b-quark masses, which shows the interface of the soft and hard QCD dynamics,
respectively. We note that the $t$ scale is chosen such that the
contributions from higher order effects are as small as possible \cite{MNP}
and one can find the specific scale to be the chiral symmetry breaking scale
\cite{KLS1,CKL}.

\item  Penguin enhancement: As known, Wilson coefficients (WCs) of $%
C_{4}(\mu )$ and $C_{6}(\mu )$ generated from the QCD penguin
\cite{BBL} increase significantly at $t<M_{B}/2$. Due to the
enhancements, it is realized that the BRs of $B\rightarrow \phi K$
in the MPQCD \cite{Mishima,CKL} can explain the results of Belle
\cite{BELLE} and BaBar \cite{BaBar} naturally.

\item  Less theoretical uncertainties: The main theoretical uncertainties in
the MPQCD from are the shape parameter $\omega _{B}$ of the B-meson wave
function, chiral symmetrical breaking parameter $m_{P}^{0}$ introduced by
the matrix element of the pseudoscalar nonlocal operator for the $P$ meson
\cite{KLS,Ball}, and the power factor $c$ for the parametrization of the
Sudakov factor $S_{t}\propto [x(1-x)]^{c}$ generated by the threshold
resummation, respectively. 
With experiments, one would precisely set the limits on these theoretical
unknown parameters.

\item  Large absorptive parts: The major absorptive parts arise from the
annihilation topology in which the propagator of the internal quark
satisfies the on-shell condition. With the power counting rule, the ratios
of the transition form factor ($F^{BK}$) to the imaginary and real parts of
annihilation contributions are found to be $F^{BK}$: $ImF^{KK}$: $ReF^{KK}$ $%
=1:2m_{K}^{0}/M_{B}$: $\Lambda /M_{B}$ \cite{CKL}. For $M_{B}\sim 5.0$ GeV,
the value of the imaginary part is compatible with that of the form factor.
\end{itemize}

In this paper, to calculate the matrix elements of four-quark operators
relevant to the $B_{s}\rightarrow KK$ decays, we adopt the MPQCD
factorization formalism as
\begin{eqnarray}
\langle VK|{\cal O}_{k}|B\rangle &=&\int [dx]\int \left[ \frac{d^{2}\vec{b}}{%
4\pi }\right] \Phi _{K}^{*}(x_{3},\vec{b}_{3})\Phi _{K}^{*}(x_{2},\vec{b}%
_{2})\,  \nonumber \\
&\times &T_{k}(\{x\},\{\vec{b}\},M_{B})\Phi _{B_{s}}(x_{1},\vec{b}_{1})
\nonumber \\
&\times &S_{t}\left( \{x\}\right) \displaystyle{e^{-\displaystyle{S\left(
\{x\},\{\vec{b}\},M_{B}\right) }}}\;  \label{eqpqcd}
\end{eqnarray}
where $\Phi _{K}^{*}$ and $\Phi _{B_{s}}$ are the wave functions of $K$ and $%
B_{s}$, $T_{k}$ is the hard scattering amplitude dictated by ${\cal O}_{k}$
operators, the exponential factor is the Sudakov factor \cite{CS,BS}, and $%
S_{t}(x)$ \cite{Li1,KLS} expresses the threshold resummation factor. Since B$%
_{s}$ is a heavy meson, the chiral symmetry breaking effects are negligible
so that $\Phi _{B_{s}}$ is regarded as $\Phi _{B_{d,u}}$.

The effective Hamiltonian for decays with the $b\to s$ transition is given
by
\begin{equation}
H_{{\rm eff}}=\frac{G_{F}}{\sqrt{2}}\sum_{q^{\prime }=u,c}V_{q^{\prime
}}\left[ C_{1}(\mu ){\cal O}_{1}^{(q^{\prime })}+C_{2}(\mu ){\cal O}%
_{2}^{(q^{\prime })}+\sum_{i=3}^{10}C_{i}(\mu ){\cal O}_{i}\right]
\end{equation}
where $V_{q^{\prime }}=V_{q^{\prime }s}^{*}V_{q^{\prime }b}$ are the
products of the CKM matrix elements, $C_{i}(\mu )$ are the WCs and ${\cal O}%
_{i}$ correspond to four-quark operators, and their detailed expressions can
be found in Ref.\cite{BBL}. It is easy to find that the structures of ${\cal %
O}_{3,4,5,6}$ , generated by the QCD penguin, are the same as that of ${\cal %
O}_{9,10,7,8}$ from the EW penguin, respectively. Except the different color
flows between $O_{2i-1}$ and $O_{2i}$ $(i=1,2,\cdots ,5)$,
each of the pair operators also has the same structure. Therefore, we can
define the useful EW dynamical variables as
\begin{eqnarray}
a_{1} &=&C_{1}+\frac{C_{2}}{N_{c}},\qquad a_{2}=C_{2}+\frac{C_{1}}{N_{c}}%
,\quad a_{1,2}^{\prime }=\frac{C_{2,1}}{N_{c}}\,,  \nonumber \\
a_{3,4}^{q} &=&C_{3,4}+\frac{3e_{q}}{2}C_{9,10}+a_{3,4}^{\prime q},\quad
a_{3,4}^{\prime q}=\frac{C_{4,3}}{N_{c}}+\frac{3e_{q}}{2N_{c}}C_{10,9}\,,
\nonumber \\
a_{5,6}^{q} &=&C_{5,6}+\frac{3e_{q}}{2}C_{7,8}+a_{5,6}^{\prime q},\quad
a_{5,6}^{\prime q}=\frac{C_{6,5}}{N_{c}}+\frac{3e_{q}}{2N_{c}}C_{8,7}\,,
\label{wcs}
\end{eqnarray}
where the superscript $q$ can be the light u, d and s quarks, respectively.
We note that all decay amplitudes related to the weak dynamical interactions
are dependent on $a_{j}^{(q)}$ and $a_{j}^{\prime (q)}$ $(j=1,2,\cdots ,6)$,
while nonfactorizable effects are only associated with the color suppressed
variables $a_{j}^{\prime (q)}$.

By defining the decay rates 
as
\[
\Gamma =\frac{G_{F}^{2}M_{B}^{3}}{32\pi }\left| {\cal A}\right| ^{2}
\]
with $G_{F}$ and $M_{B}$ being the Fermi constant and the mass of $B_{s}$,
the corresponding decay amplitudes for $B_{s}\rightarrow K^{+}K^{-}$ and $%
B_{s}\rightarrow K^{0}\bar{K}^{0}$ modes are written by
\begin{eqnarray}
{\cal A}^{+-} &=&f_{K}V_{t}^{*}F_{e46}^{P(u)}+V_{t}^{*}{\cal M}%
_{e46}^{P(u)}+f_{B}V_{t}^{*}F_{a6}^{P(s)}+V_{t}^{*}({\cal M}_{a46}^{P(s)}+%
{\cal M}_{a35}^{P(s)}+{\cal M}_{a35}^{P(u)})  \nonumber \\
&&-f_{K}V_{u}^{*}F_{e2}-V_{u}^{*}{\cal M}_{e2}-V_{u}^{*}{\cal M}_{a1},
\nonumber \\
{\cal A}^{00} &=&f_{K}V_{t}^{*}F_{e46}^{P(d)}+V_{t}^{*}{\cal M}%
_{e46}^{P(d)}+f_{B}V_{t}^{*}F_{a6}^{P(s)}+V_{t}^{*}({\cal M}_{a46}^{P(s)}+%
{\cal M}_{a35}^{P(s)}+{\cal M}_{a35}^{P(d)})\,,  \label{namp}
\end{eqnarray}
respectively, where $f_{K(B)}$ is the $K(B)$ decay constant. In Eq. (\ref
{namp}), $\{F_{e(a)}^{[P(q]}\}$ denote the factorizable emission
(annihilation) transition form factors for tree [penguin] diagram
contributions, while $\{{\cal M}_{e(a)}^{[P(q)]}\}$ correspond to
nonfactorizable effects, the superscript $q$ represents the $q$ quark pair
emitted from the EW penguins, the subscripts of 1-6 label the WCs of Eq. (%
\ref{wcs}) appearing in the factorization formulas, and the nonfactorizable
amplitudes ${\cal M}_{e2(a1)}$ are from the operators $O_{1,2}^{(u)}$. Note
that there is no tree contribution to the decay of $B_{s}\rightarrow K^{0}%
\bar{K}^{0}$. Explicitly, one has that
\begin{eqnarray}
F_{e46}^{P(q)} &=&F_{4}^{P(q)}+F_{6}^{P(q)}\;,  \nonumber \\
F_{4}^{P(q)} &=&8\pi C_{F}M_{B}^{2}\int_{0}^{1}dx_{1}dx_{2}\int_{0}^{\infty
}b_{1}db_{1}b_{2}db_{2}\phi _{B}(x_{1},b_{1})  \nonumber \\
&&\times \left\{ \left[ (1+x_{2})\phi _{K}(1-x_{2})+r_{K}(1-2x_{2})\left(
\phi _{K}^{p}(x_{2})+\phi _{K}^{s}(x_{2})\right) \right] \right.  \nonumber
\\
&&\times E_{e4}^{(q)}(t_{e}^{(1)})h_{e}(x_{1},x_{2},b_{1},b_{2})  \nonumber
\\
&&\left. +2r_{K}\phi
_{K}^{p}(x_{2})E_{e4}^{(q)}(t_{e}^{(2)})h_{e}(x_{2},x_{1},b_{2},b_{1})\right%
\} \;,  \label{int4} \\
F_{6}^{P(q)} &=&16\pi
C_{F}M_{B}^{2}r_{K}\int_{0}^{1}dx_{1}dx_{2}\int_{0}^{\infty
}b_{1}db_{1}b_{2}db_{2}\phi _{B}(x_{1},b_{1})  \nonumber \\
&&\times \left\{ \left[ \phi _{K}(1-x_{2})+r_{K}\left( (2+x_{2})\phi
_{K}^{p}(x_{2})-x_{2}\phi _{K}^{s}(x_{2})\right) \right] \right.  \nonumber
\\
&&\times E_{e6}^{(q)}(t_{e}^{(1)})h_{e}(x_{1},x_{2},b_{1},b_{2})  \nonumber
\\
&&\left. +\left[ 2r_{K}\phi _{K}^{p}(x_{2})\right]
E_{e6}^{(q)}(t_{e}^{(2)})h_{e}(x_{2},x_{1},b_{2},b_{1})\right\} \;,
\label{int6} \\
F_{a6}^{P(s)} &=&16\pi
C_{F}M_{B}^{2}r_{K}\int_{0}^{1}dx_{2}dx_{3}\int_{0}^{\infty
}b_{2}db_{2}b_{3}db_{3}  \nonumber \\
&&\times \left\{ \left[ 2\phi _{K}^{p}(x_{2})\phi _{K}(x_{3})+x_{3}\phi
_{K}(x_{2})\left( \phi _{K}^{p}(x_{3})-\phi _{K}^{s}(x_{3})\right) \right]
\right.  \nonumber \\
&&\times E_{a6}^{(s)}(t_{a}^{(1)})h_{a}(x_{2},x_{3},b_{2},b_{3})  \nonumber
\\
&&+\left[ 2\phi _{K}(x_{2})\phi _{K}^{p}(x_{3})+x_{2}\phi _{K}(x_{3})\left(
\phi _{K}^{p}(x_{2})-\phi _{K}^{s}(x_{2})\right) \right]  \nonumber \\
&&\left. \times
E_{a6}^{(s)}(t_{a}^{(2)})h_{a}(x_{3},x_{2},b_{3},b_{2})\right\} \;,
\label{exc6}
\end{eqnarray}
with $r_{K}=m_{K}^{0}/M_{B}$ for $q=u$- or $d$-quark. Here, $C_{F}=4/3$ is
the color factor, $\phi _{B}$ and $\phi _{K}$ belong to twist-2 $B_{s}$ and $%
K$ wave functions, while $\phi _{K}^{p}$ and $\phi _{K}^{s}$ are for twist-3
\cite{Ball} and their detailed expressions can be found in Ref. \cite{CKL},
the hard part functions $h_{e(a)}$ have included the $S_{t}$ factor \cite
{CKL} and the evolution factors are given by
\begin{eqnarray}
E_{ei}^{(q)}(t) &=&\alpha _{s}(t)a_{i}^{(q)}(t)\exp
[-S_{B}(t,x_{1})-S_{K}(t,x_{2})]\;,  \label{e1} \\
E_{ai}^{(q)}(t) &=&\alpha _{s}(t)a_{i}^{(q)}(t)\exp
[-S_{K}(t,x_{2})-S_{K}(t,x_{3})]\,,
\end{eqnarray}
respectively. As expected that the nonfactorizable effects are smaller and
more complicated, we do not display their expressions here but they will be
included in the numerical calculations. We note that in our PQCD approach
the $x$ dependence in the kaon wave function is usually assigned to the
momentum fraction of the light $u$ or $d$ quark \cite{CKL}. However, due to
the $s$ quark being spectator in emission contributions, for simplified the
hard parts, we adopt that this $s$ quark carries a momentum fraction of $x$
so that in order to compensate this change, the arguments $x_{2}$ of the
kaon wave functions $\left\{ \phi _{K}\left( x_{2}\right) \right\} $ in
Eqs.~(\ref{int4}) and (\ref{int6}) are replaced by $1-x_{2}$ in which we
also have used the properties of $\phi _{K}^{p}\left( 1-x_{2}\right) =\phi
_{K}^{p}\left( x_{2}\right) $ and $\phi _{K}^{s}\left( 1-x_{2}\right) =-\phi
_{K}^{s}\left( x_{2}\right) $ \cite{Ball,CKL}. On the other hand, the
factorizable annihilation contribution, associated with WC $a_{4}^{(q)}$ and
described by
\begin{eqnarray}
F_{a4}^{P(q)} &=&16\pi
C_{F}M_{B}^{2}\int_{0}^{1}dx_{2}dx_{3}\int_{0}^{\infty
}b_{2}db_{2}b_{3}db_{3}  \nonumber \\
&&\times \left\{ \left[ x_{3}\phi _{K}(x_{2})\phi _{K}(x_{3})+2r_{K}^{2}\phi
_{K}^{p}(x_{2})\left( (1+x_{3})\phi _{K}^{p}(x_{3})\right. \right. \right.
\nonumber \\
&&\hspace{1cm}\left. \left. -(1-x_{3})\phi _{K}^{s}(x_{3})\right) \right]
E_{a4}^{(q)}(t_{a}^{(1)})h_{a}(x_{2},x_{3},b_{2},b_{3})  \nonumber \\
&&-\left[ x_{2}\phi _{K}(x_{2})\phi _{K}(x_{3})+2r_{K}^{2}\phi
_{K}^{p}(x_{3})\left( (1+x_{2})\phi _{K}^{p}(x_{2})\right. \right.  \nonumber
\\
&&\hspace{1cm}\left. \left. \left. -(1-x_{2})\phi _{K}^{s}(x_{2})\right)
\right] E_{a4}^{(q)}(t_{a}^{(2)})h_{a}(x_{3},x_{2},b_{3},b_{2})\right\} \;,
\label{exc4}
\end{eqnarray}
vanishes. It can be seen easily by interchanging the integration variables $%
x_{2}$ and $x_{3}$ in the second terms of Eq.(\ref{exc4}). However, this
property does not apply to the annihilation contributions associated with $%
a_{6}^{(q)}$ which are constructive in Eq.~(\ref{exc6}). The factorization
formulas for $F_{a1}$ and $F_{a35}^{P(q)}$, associated with the WCs of $%
a_{1}(t_{a})$ and $a_{3}^{(q)}(t_{a})+a_{5}^{(q)}(t_{a})$, are the same as $%
F_{a4}^{P(s)}$, {\it i.e.}, vanish.

\begin{table}[h]
\caption{ The values of individual transition form factors (TFFs) for $%
B_{s}\rightarrow KK$ decays with $\omega_{B}=0.4$, two sets of $m^0_{K}$ and
${\cal M}^{P(q)}_{aP}={\cal M}^{P(q)}_{a35}+{\cal M}^{P(s)}_{a35}+{\cal M}%
^{P(s)}_{a46}$.}
\label{ff}
\begin{center}
$
\begin{tabular}{|c|c|c|c|c|c|}
\hline
{\rm TFF} & $F_{e46}^{P(u)}$ & $F_{e46}^{P(d)}$ & $F_{a6}^{P(s)}$ & $F_{e2}$
& ${\cal M}_{e46}^{P(u)}$ \\ \hline
$m_{K}^{0}=1.7$ & $-31.91$ & $-32.67$ & $-1.25+i10.91$ & $369.37$ & $%
0.17+i0.32$ \\ \hline
$m_{K}^{0}=1.5$ & $-27.00$ & $-27.64$ & $-1.10+i9.63$ & $336.53$ & $%
0.16+i0.28$ \\ \hline\hline
{\rm TFF} & ${\cal M}_{e46}^{P(d)}$ & ${\cal M}_{aP}^{P(u)}$ & ${\cal M}%
_{aP}^{P(d)}$ & ${\cal M}_{e2}$ & ${\cal M}_{a1}$ \\ \hline
$m_{K}^{0}=1.7$ & $0.16+i0.48$ & $0.19+i0.52$ & $0.20+i0.54$ & $-1.19-i2.75$
& $0.69-i4.3$ \\ \hline
$m_{K}^{0}=1.5$ & $0.16+i0.45$ & $0.15+i0.50$ & $0.15+i0.52$ & $-1.36-i2.95$
& $0.59-i4.04$ \\ \hline
\end{tabular}
$%
\end{center}
\end{table}

\begin{figure}[tbp]
\hspace{2.5cm} \psfig{figure=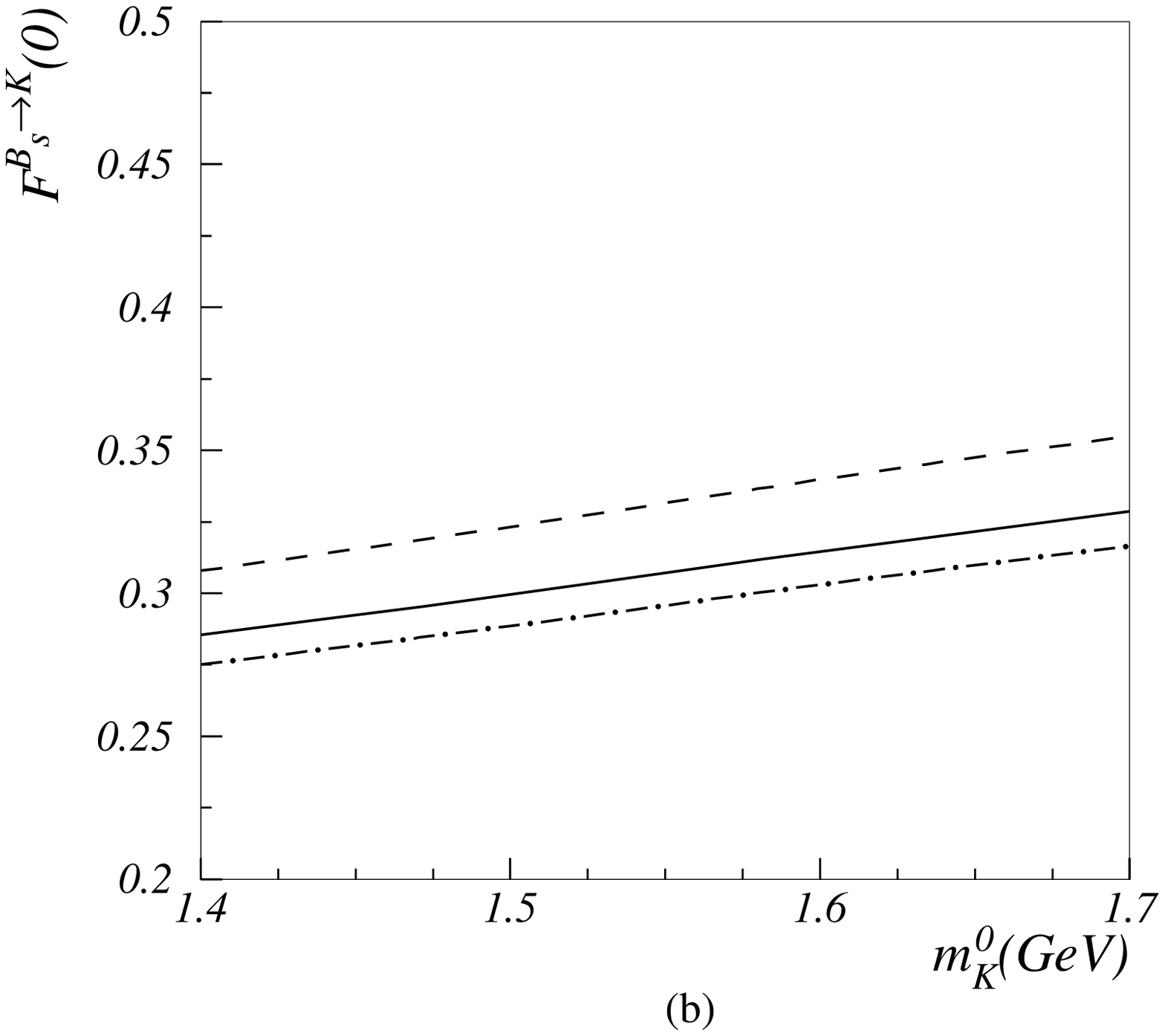,height=3.5in }
\caption{Form factors of (a) $B_{d}\rightarrow K$ and (b) $B_{s}\rightarrow
K $ with respect to $m^0_{K}$. The dashed, solid and dashed-dot lines
correspond to $\omega_{B}=$0.38, 0.4 and 0.41 GeV, respectively.}
\label{ffbsk}
\end{figure}

In the numerical analysis, we first show the $B_{s}\rightarrow K$ form
factor of $F^{B_{s}K}$ in Figure \ref{ffbsk}, which can be easily obtained
by dropping the WC dependence out of Eq. (\ref{int4}),
where we have used $f_{B}=0.20$ and $f_{K}=0.16$ GeV. For a comparison, we
also display the results of $B_{d}\rightarrow K$ form factor ($F^{B_{d}K}$)
from $\left\{ \phi _{K}\left( x_{2}\right) \right\} $ and $M_{B_{d}}$
instead of $\left\{ \phi _{K}\left( 1-x_{2}\right) \right\} $ and $M_{B_{s}}$%
, respectively. From the figure, we see that $F^{B_{s}K}$ is slightly less
than $F^{B_{d}K}$ and the main difference is from the chiral symmetry
breaking effect appearing in $\phi _{K}$ arisen from the different argument
of $\phi _{K}$. After including WCs, the values of individual terms in Eq. (%
\ref{namp}) can be read from Table \ref{ff}. The slight difference between $%
F_{e46}^{P(u)}$ and $F_{e46}^{P(d)}$ comes from the EW effects. The
factorizable annihilation and nonfactorizable contributions are complex
because the on-shell condition can only be satisfied in these diagrams. On
the other hand, in order to know the strong phase more clearly, we
reparametrize the decay amplitudes as
\begin{equation}
A=V_{t}^{*}P-V_{u}^{*}T=-V_{c}^{*}P\left( 1+\left| \frac{V_{u}}{V_{c}}%
\right| re^{i(\delta +\phi _{3})}\right)  \label{amp}
\end{equation}
with $re^{i\delta }=1+T/P$, where we have used
$\sum_{i=\{u,c,t\}}V_{i}=0$, $T$ and $P$ express the whole tree
and penguin contributions, $\delta $ describes the
strong phase, and the values of $r$ and $\delta $ are shown in Table \ref{sp}%
. From the table, we clearly see that cos$\delta <0$ in the MPQCD approach.

\begin{table}[h]
\caption{ The results of $r$ and $\delta$ with different values of $%
\omega_{B}$ and $m^0_{K}$.}
\label{sp}
\begin{center}
\begin{tabular}{|l|ll|ll|ll|}
\hline
$m_{K}^{0}$ &  & $1.7$ &  & $1.6$ &  & $1.5$ \\ \hline\hline
\multicolumn{1}{|c|}{$\omega _{B}$} & \multicolumn{1}{|c}{$r$} &
\multicolumn{1}{|c|}{$\delta (\deg )$} & \multicolumn{1}{|c}{$r$} &
\multicolumn{1}{|c|}{$\delta (\deg )$} & \multicolumn{1}{|c}{$r$} &
\multicolumn{1}{|c|}{$\delta (\deg )$} \\ \hline
\multicolumn{1}{|c|}{$0.41$} & \multicolumn{1}{|c}{$9.16$} &
\multicolumn{1}{|c|}{$207.7$} & \multicolumn{1}{|c}{$9.42$} &
\multicolumn{1}{|c|}{$208.0$} & \multicolumn{1}{|c}{$9.70$} &
\multicolumn{1}{|c|}{$208.3$} \\ \hline
\multicolumn{1}{|c|}{$0.40$} & \multicolumn{1}{|c}{$9.22$} &
\multicolumn{1}{|c|}{$206.9$} & \multicolumn{1}{|c}{$9.49$} &
\multicolumn{1}{|c|}{$207.2$} & \multicolumn{1}{|c}{$9.78$} &
\multicolumn{1}{|c|}{$207.6$} \\ \hline
\multicolumn{1}{|c|}{$0.38$} & \multicolumn{1}{|c}{$9.34$} &
\multicolumn{1}{|c|}{$205.3$} & \multicolumn{1}{|c}{$9.63$} &
\multicolumn{1}{|c|}{$205.6$} & \multicolumn{1}{|c}{$9.92$} &
\multicolumn{1}{|c|}{$205.9$} \\ \hline
\end{tabular}
\end{center}
\end{table}

\begin{figure}[tbp]
\hspace{2.5cm} \psfig{figure=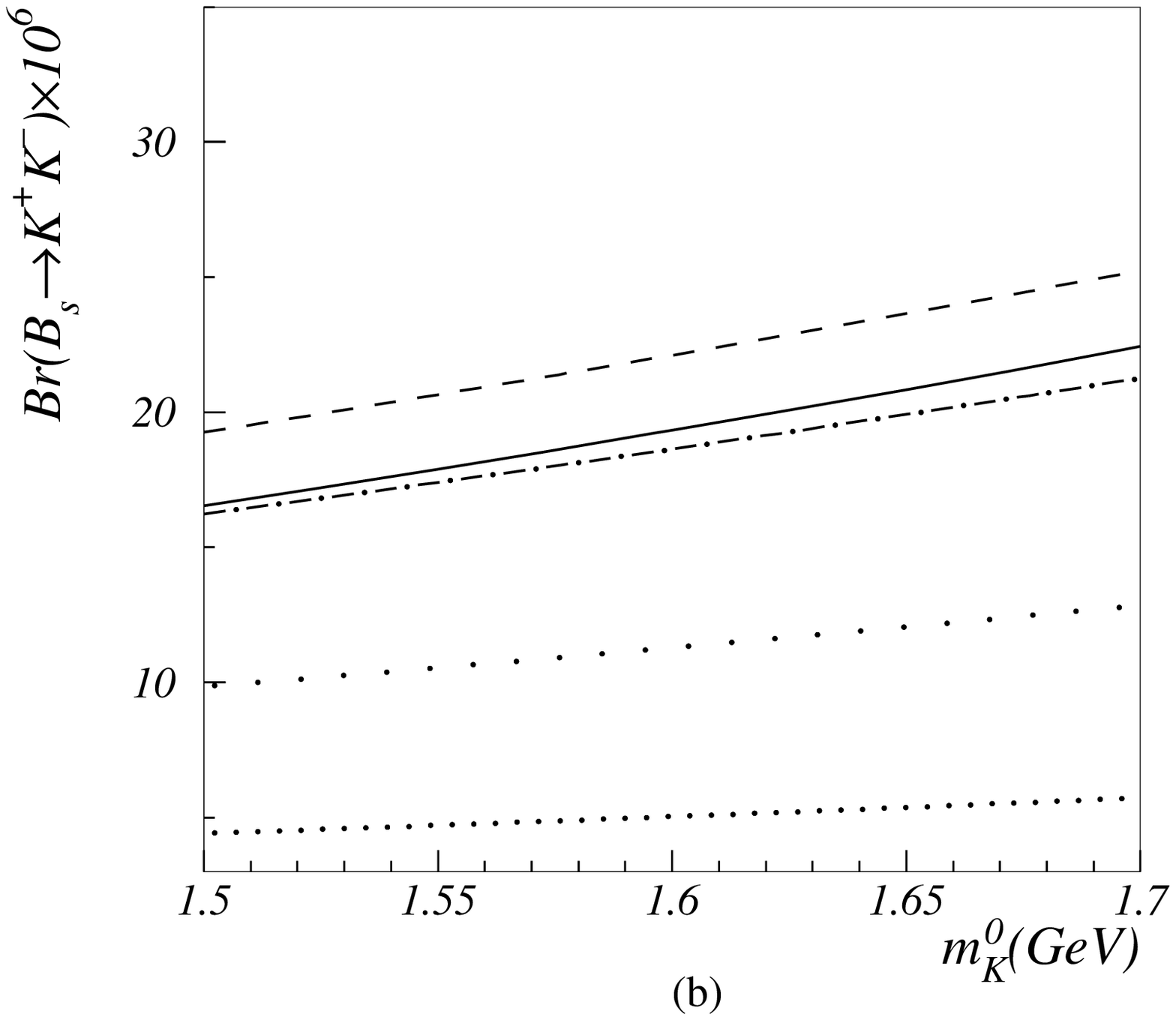,height=3.5in }
\caption{ BRs (in units of $10^{-6}$) of (a) $B_s\to K^0 \bar{K}^0$ and (b) $%
B_s\to K^+K^-$ with fixing the typical scale on WCs. The dashed, dashed-dot,
dotted and dense-dot lines stand for t=1.5, 1.7, 2.5, 4.8 GeV, with $%
\omega_{B}=0.4$ GeV and $\phi_{3}=72^0$ while the solid line is the result
of $t$ as a running scale, respectively.}
\label{scale}
\end{figure}

\begin{figure}[tbp]
\hspace{2.5cm} \psfig{figure=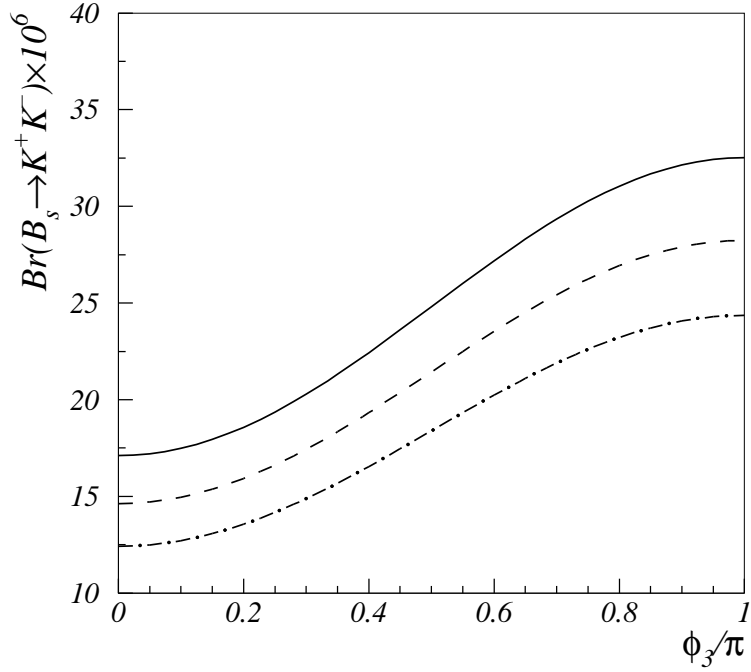,height=3.5in }
\caption{ The CP average BR of $B_{s}\rightarrow K^+ K^-$ as a function of $%
\phi_{3}$ with $\omega_{B}=0.4$ GeV. The solid, dashed and dashed-dot lines
denote $m^0_{K}=$1.7, 1.6 and 1.5 GeV, respectively.}
\label{figbrang}
\end{figure}

To calculate the CP average BRs of $B_{s}\rightarrow KK$ decays, we use the
following data as input values:
\[
V_{us}\approx \lambda ,\ V_{ts}\approx -A\lambda ^{2},\ V_{ub}\approx
A\lambda ^{3}R_{b}e^{-i\phi _{3}},\ A\approx 0.80,\ \lambda \approx 0.22,\
R_{b}\approx 0.36.
\]
From these values, the results with possible $\omega _{B}$ and $m_{K}^{0}$
are displayed in Table \ref{br}. We note that although the transition form
factor of the tree contribution is larger than others by over one order of
the magnitude as shown in Table \ref{ff}, due to the suppression of CKM
matrix elements, actually the tree contribution is subdominant. By fixing
WCs at some specific scales, the results are given in Figure \ref{scale}.
From that, we clearly see that the typical scale for the $B_{s}\rightarrow
KK $ decays is around $1.7$ GeV. In Figure \ref{figbrang}, we also show the
BR of $B_s\rightarrow K^{+}K^{-}$ as a function of angle $\phi _{3}$.

\begin{figure}[tbp]
\hspace{2.5cm} \psfig{figure=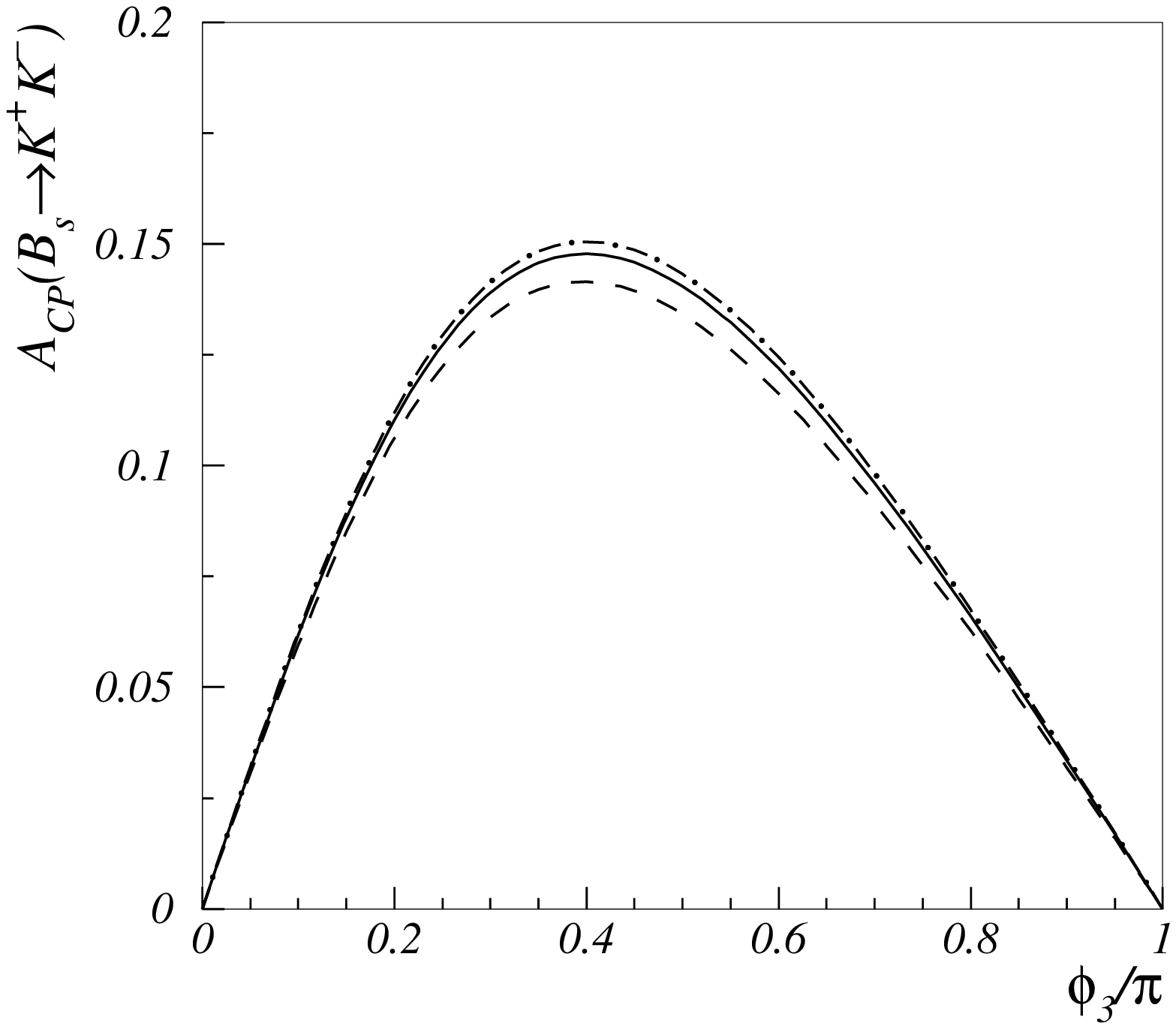,height=3.5in }
\caption{ The CP asymmetry of $B_{s}\rightarrow K^+ K^-$ with $m^0_K=1.7$
GeV. The dashed, solid and dashed-dot lines represent $\omega_{B}=$0.41, 0.4
and 0.38 GeV, respectively.}
\label{figcp}
\end{figure}

\begin{table}[tbp]
\caption{The BRs (in units of $10^{-6}$) of $B_{s}\rightarrow KK$ decays
with different values of $\omega _{B}$ and $m_{K}^{0}$. }
\label{br}
\begin{center}
$
\begin{tabular}{|c|ccc||ccc|}
\hline
&  & $B_{s}\rightarrow K^{+}K^{-}$ &  &  & $B_{s}\rightarrow K^{0}\bar{K}%
^{0} $ &  \\ \cline{2-7}
$Br$ & $m_{K}^{0}=1.7$ & \multicolumn{1}{|c}{$m_{K}^{0}=1.6$} &
\multicolumn{1}{|c||}{$m_{K}^{0}=1.5$} & $m_{K}^{0}=1.7$ &
\multicolumn{1}{|c}{$m_{K}^{0}=1.6$} & \multicolumn{1}{|c|}{$m_{K}^{0}=1.5$}
\\ \hline
$\omega _{B}=0.41$ & $21.08$ & \multicolumn{1}{|c}{$18.18$} &
\multicolumn{1}{|c||}{$15.58$} & $24.22$ & \multicolumn{1}{|c}{$20.96$} &
\multicolumn{1}{|c|}{$18.04$} \\ \hline
$\omega _{B}=0.40$ & $22.43$ & \multicolumn{1}{|c}{$19.33$} &
\multicolumn{1}{|c||}{$16.55$} & $25.78$ & \multicolumn{1}{|c}{$22.32$} &
\multicolumn{1}{|c|}{$19.20$} \\ \hline
$\omega _{B}=0.38$ & $25.56$ & \multicolumn{1}{|c}{$21.97$} &
\multicolumn{1}{|c||}{$18.77$} & $29.50$ & \multicolumn{1}{|c}{$25.45$} &
\multicolumn{1}{|c|}{$21.84$} \\ \hline
\end{tabular}
$%
\end{center}
\end{table}
As usual, the direct CP\ asymmetry (CPA) can be defined as
\begin{eqnarray}
A_{CP} &=&\frac{\Gamma (B_{s}\rightarrow K^{+}K^{-})-\Gamma (\bar{B}%
_{s}\rightarrow K^{-}K^{+})}{\Gamma (B_{s}\rightarrow K^{+}K^{-})+\Gamma (%
\bar{B}_{s}\rightarrow K^{-}K^{+})},  \nonumber \\
&=&-\frac{2\lambda ^{2}R_{b}rsin\delta sin\phi _{3}}{\left( 1+\left( \lambda
^{2}R_{b}r\right) ^{2}{\large +}2\lambda ^{2}R_{b}rcos\delta cos\phi
_{3}\right) }  \label{cpa}
\end{eqnarray}
by using Eq. (\ref{amp}). From Eq. (\ref{cpa}), we see that the CPA is
associated with both weak CP and strong phases. The results as a function of
angle $\phi _{3}$ are plotted in Figure \ref{figcp}.

In the following, we present the implications of our results. With the limit
of SU(3) symmetry, it is known that $\Gamma (B_{d}\rightarrow K^{+}\pi
^{-})\approx \Gamma (B_{s}\rightarrow K^{+}K^{-})$ and $\Gamma
(B_{d}\rightarrow K^{0}\bar{K}^{0})\approx \left| V_{td}/V_{ts}\right|
^{2}\Gamma (B_{s}\rightarrow K^{+}K^{-})$ \cite{GHLR}. With including SU(3)
breaking effects, one has \cite{CQ}
\begin{eqnarray}
Br\left( B_{s}\rightarrow K^{\pm }K^{\mp }\right) &\approx &\frac{\tau
_{B_{s}}}{\tau _{B_{d}}}\left( \frac{M_{B_{s}}}{M_{B_{d}}}\right) ^{3}\left(
\frac{F^{B_{s}K}(0)}{F^{B_{d}\pi }(0)}\right) ^{2}Br\left( B_{d}\rightarrow
K^{\pm }\pi ^{\mp }\right) ,  \label{br1} \\
Br\left( B_{s}\rightarrow K^{0}\bar{K}^{0}\right) &\approx &\frac{\tau
_{B_{s}}}{\tau _{B_{d}}}\left( \frac{M_{B_{s}}}{M_{B_{d}}}\right) ^{3}\left|
\frac{V_{ts}}{V_{td}}\frac{F^{B_{s}K}(0)}{F^{B_{d}K}(0)}\right| ^{2}Br\left(
B_{d}\rightarrow K^{0}\bar{K}^{0}\right) .  \label{br2}
\end{eqnarray}
We note that $Br\left( B_{d}\rightarrow K^{\pm }\pi ^{\mp }\right)
$ has been observed in the present B factories
\cite{BCP41,BCP42,pipiCLEO}
and although there is no limit on $Br\left( B_{d}\rightarrow K^{0}\bar{K}%
^{0}\right) $, its estimation has been done by the PQCD in Ref. \cite{CL}.
By taking $Br\left( B_{d}\rightarrow K^{\pm }\pi ^{\mp }\right) \simeq
18.5\times 10^{-6}$ and $Br\left( B_{d}\rightarrow K^{0}\bar{K}^{0}\right)
\simeq 1.4\times 10^{-6}$, we get $Br\left( B_{s}\rightarrow K^{\pm }K^{\mp
}\right) \simeq 22.68\times 10^{-6}$ and $Br\left( B_{s}\rightarrow K^{0}%
\bar{K}^{0}\right) \simeq 26.05\times 10^{-6}$. Comparing with the
values in Table \ref{br}, obviously the results are consistent
with the those from the MPQCD for $\omega _{B}=0.4,$
$m_{K}^{0}=1.7$ GeV and $\phi _{3}\simeq 72^{0}. $ If so, the
MPQCD prefers $\phi _{3}$ to be less than $90^{0}.$ On the other
hand, if neglecting the small difference from EW effects ($\sim
7\%$ difference in our analysis), from Eq. (\ref{amp}), we find
that \cite {CQ}
\begin{equation}
Br\left( B_{s}\rightarrow K^{\pm }K^{\mp }\right) \approx Br\left(
B_{s}\rightarrow K^{0}\bar{K}^{0}\right) \left( 1+2\lambda ^{2}R_{b}r\cos
\delta \cos \phi _{3}\right) .  \label{br3}
\end{equation}
In Eq. (\ref{br3}), if the interference term associated with cos$\phi _{3}$%
cos$\delta $ is negative, it gives $Br(B_{s}\rightarrow $ $K^{\pm
}K^{\mp })$$<$ $Br(B_{s}\rightarrow $ $K^{0}\bar{K}^{0})$. In the
MPQCD, it also implies
that $\phi _{3}<90^0$. From Eq. (\ref{br1}), we also expect that $%
A_{CP}\left( B_{s}\rightarrow K^{\pm }K^{\mp }\right) \approx A_{CP}\left(
B_{d}\rightarrow K^{\pm }\pi ^{\mp }\right) $. Hence, by the measurements of
$B_{s}\rightarrow KK$ decays, the sign of cos$\phi _{3}$cos$\delta $ can be
determined and the predictions of the MPQCD can also be verified.

Finally, we give a brief remark on the relations between $B_{s}\rightarrow
K^{\pm }K^{\mp }$ and $B_{d}\rightarrow \pi ^{\pm }\pi ^{\mp }$ decays. It
is known that, as shown in Ref. \cite{Fleischer2}, there are close
relationships for transition form factors between the decays by interchange
of $d$ and $s$ quarks, called the U-spin transformation. Under the U-spin
limit, both decays approximately have the same $r$ and $\delta $ as defined
in Eq. (\ref{amp}). Moreover, we find that the term associated with $\cos
\delta \cos \phi _{3}$ in the $B_{d}\rightarrow \pi ^{\pm }\pi ^{\mp }$
decay has an opposite sign to that in $B_{s}\rightarrow K^{\pm }K^{\mp }$.
This implies that while one of both distributions increases with respect to $%
\phi _{3}$, the other one will decrease \cite{CQ}. Hence, by precise
measurements on the BRs of $B_{d}\rightarrow \pi ^{\pm }\pi ^{\mp }$ and $%
B_{s}\rightarrow K^{\pm }K^{\mp }$, we also can obtain the information of $%
\cos \delta \cos \phi _{3}$.

In summary, we have studied the $B_{s}\rightarrow KK$ decays with the MPQCD
approach. We have verified that the typical scale in the MPQCD is around $%
1.7 $ GeV and the large absorptive parts make $sin\delta \sim -0.45$ so that
the CP asymmetry could be as large as $15\%$. Since the Tevatron Run II has
started a new physics run at $\sqrt{s}=2$ ${\rm TeV}$ and will collect a
data sample of $2$ {\rm fb}$^{-1}$ in the first two years \cite{CDF1}. At
its initial phase with 10K of $B_{s}\rightarrow KK$ events, the BRs and CP
asymmetry predicted by the MPQCD can be tested.\newline

\noindent {\bf Acknowledgments}

The author would like to thank C. Q. Geng, Y.Y. Keum and H.N. Li
for their useful discussions. This work was supported in part by
the National Science Council of the Republic of China under
contract numbers NSC-89-2112-M-006-033 and the National Center for
Theoretical Science.


\newpage

\begin{figcap}
\item
  Form factors of (a) $B_{d}\rightarrow K$  and (b) $B_{s}\rightarrow K$
 with respect to $m^0_{K}$. The dashed, solid and
dashed-dot lines correspond to $\omega_{B}=$0.38, 0.4 and 0.41
GeV, respectively.
\item
 BRs (in units of $10^{-6}$)  of (a) $B_s\to K^0 \bar{K}^0$ and (b)
$B_s\to K^+K^-$  with fixing the typical scale on WCs. The dashed,
dashed-dot, dotted and dense-dot lines stand for t=1.5, 1.7, 2.5,
4.8 GeV, with $\omega_{B}=0.4$ GeV and $\phi_{3}=72^0$
while the solid line is the result of $t$ as a running scale,
respectively.
\item
The CP average BR of $B_{s}\rightarrow K^+ K^-$ as a function of
$\phi_{3}$ with $\omega_{B}=0.4$ GeV. The solid, dashed and
dashed-dot lines denote  $m^0_{K}=$1.7, 1.6 and 1.5
GeV, respectively.
\item
The CP asymmetry of $B_{s}\rightarrow K^+ K^-$  with $m^0_K=1.7$
GeV. The dashed, solid and dashed-dot lines represent
 $\omega_{B}=$0.41, 0.4 and 0.38 GeV, respectively.
\end{figcap}

\newpage
\begin{figure}[tbp]
\hspace{2.5cm} \psfig{figure=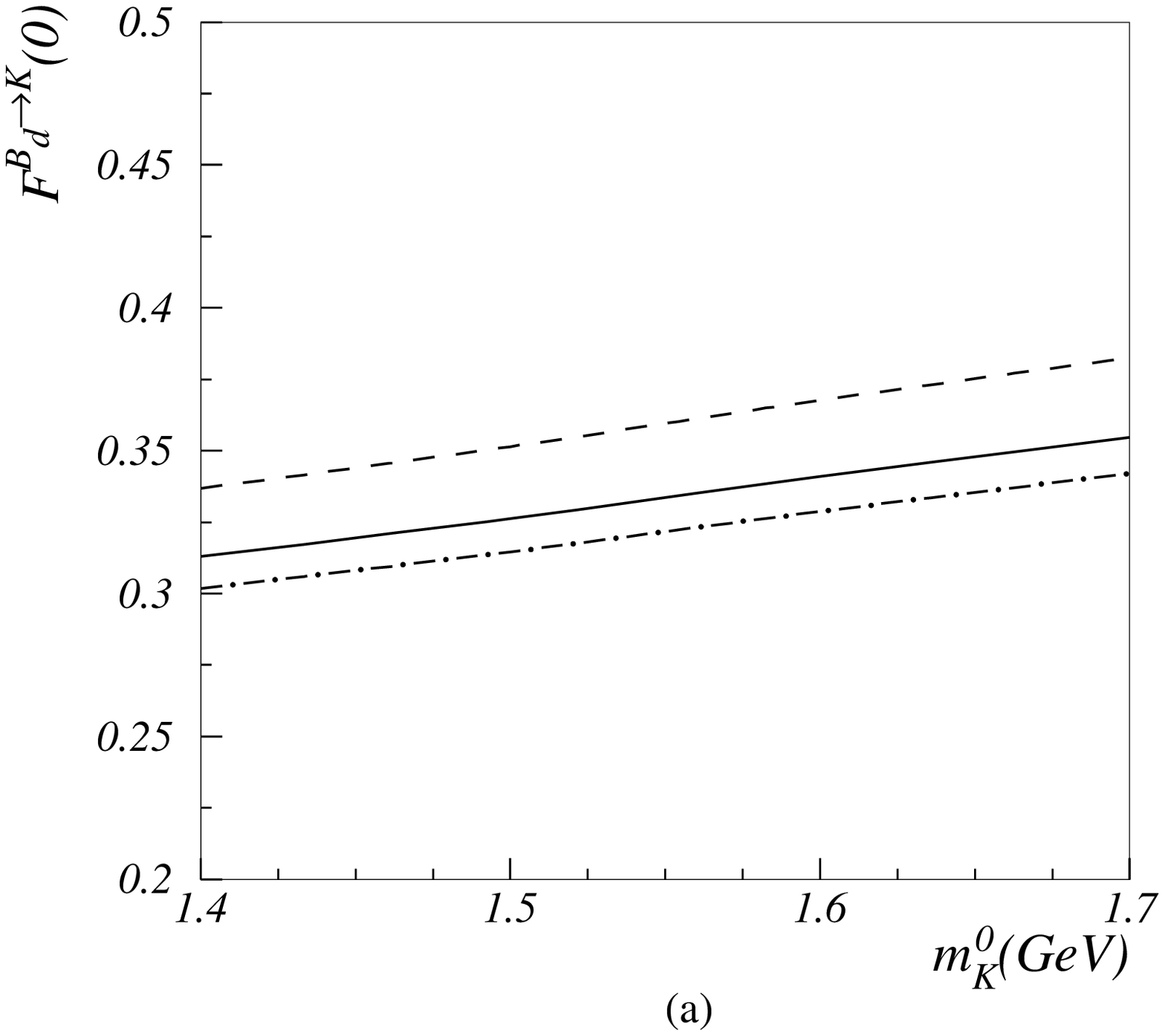,height=3.5in }
\end{figure}

\newpage
\begin{figure}[tbp]
\hspace{2.5cm} \psfig{figure=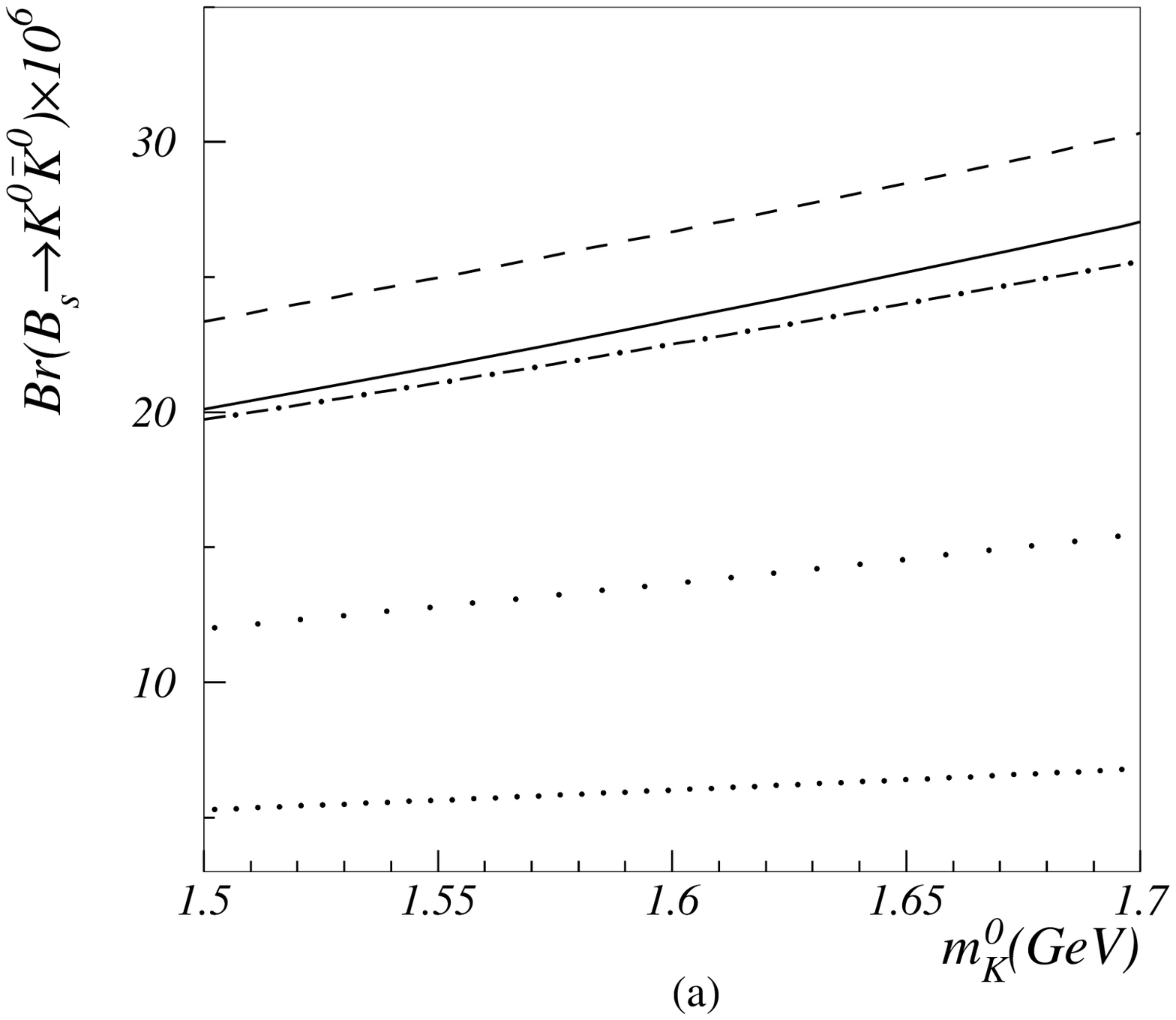,height=3.5in }
\end{figure}

\newpage

\newpage

\begin{thebibliography}{99}
\bibitem{CLEO}  CLEO Collaboration, T. Bergfeld {\it et al.}, Phys. Rev.
Lett. {\bf 81}, 272 (1998).

\bibitem{BCP41}  BABAR Collaboration, B. Aubert {\it et al., }%
hep-ex/0105061.

\bibitem{BCP42}  Belle Collaboration, K. Abe {\it et al}.,
hep-ex/0104030.

\bibitem{CKM}  M. Kobayashi and T. Maskawa, Prog. Theor. Phys. {\bf 49}, 652
(1973).

\bibitem{Fleischer2}  R. Fleischer, Phys. Lett. B{\bf 459}, 306 (1999).

\bibitem{CQ}  C.H. Chen and C.Q. Geng, hep-ph/0107145.

\bibitem{LB1}  G.P. Lepage and S.J. Brodsky, Phys. Lett. B{\bf 87}, 359
(1979); Phys. Rev. D{\bf 22}, 2157 (1980).

\bibitem{IL}  N. Isgur and C.H. Llewellyn-Smith, Nucl. Phys. B{\bf 317}, 526
(1989).

\bibitem{SHB}  A. Szczepaniak, E. M. Henley, and S. Brodsky, Phys. Lett. B%
{\bf 243}, 287 (1990).

\bibitem{ASY}  R. Akhoury, G. Sterman and Y.P. Yao, Phys. Rev. D{\bf 50},
358 (1994).


\bibitem{BF}  M. Beneke and T. Feldmann, Nucl. Phys. B{\bf 592}, 3 (2000).

\bibitem{KR}  A. Khodjamirian and R. Ruckl, Phys. Rev. D{\bf 58}, 054013
(1998).

\bibitem{LS}  H.N. Li and G. Sterman, Nucl. Phys. B{\bf 381}, 129 (1992).

\bibitem{S0}  G. Sterman, Phys. Lett. B{\bf 179}, 281 (1986); Nucl. Phys. B%
{\bf 281}, 310 (1987).

\bibitem{CT}  S. Catani and L. Trentadue, Nucl. Phys. B{\bf 327}, 323
(1989); Nucl. Phys. B{\bf 353}, 183 (1991).


\bibitem{CS}  J.C. Collins and D.E. Soper, Nucl. Phys. B{\bf 193}, 381
(1981).

\bibitem{BS}  J. Botts and G. Sterman, Nucl. Phys. B{\bf 325}, 62 (1989).
{\large {\bf \ }}

\bibitem{Li1}  H.N. Li, hep-ph/0102013.

\bibitem{KLS}  T. Kurimoto, H.N. Li, and A.I. Sanda, hep-ph/0105003, to appear in PRD.

\bibitem{MNP}  B. Meli\'{c}, B. Ni\v{z}i\'{c} and K. Passek, Phys. Rev. D%
{\bf 60}, 074004 (1999).

\bibitem{KLS1}  Y.Y. Keum, H.N. Li, and A.I. Sanda, Phys. Lett. B{\bf 504},
6 (2001); Phys. Rev. D{\bf 63}, 054008 (2001).

\bibitem{MPQCD} H.N. Li, Phys. Rev. D{\bf 64}, 014019 (2001).

\bibitem{pipi}  C.D. L${\rm \ddot{u}}$, K. Ukai, and M.Z. Yang, Phys. Rev. D%
{\bf 63}, 074009 (2001).

\bibitem{CL}  C.H. Chen and H.N. Li, Phys. Rev. D{\bf 63}, 014003 (2001).

\bibitem{KS}  E. Kou and A.I. Sanda, hep-ph/0106159.

\bibitem{Melic}  B. Melic, Phys. Rev. D{\bf 59}, 074005 (1999).

\bibitem{LY}  C.D. L${\rm \ddot{u}}$ and MZ. Yang, hep-ph/0011238.

\bibitem{Mishima} S. Mishima, hep-ph/0107163.

\bibitem{CKL}  C.H. Chen, Y.Y. Keum and H.N. Li, hep-ph/0107165, to appear in PRD.

\bibitem{BELLE}  Belle Collaboration, A. Bozek, talk presented at the 4th
International Workshop on B Physics and CP Violation, Ise-Shima, Japan, Feb.
19-23, 2001.

\bibitem{BaBar}  BaBar Collaboration, G. Cavoto, talk presented at the XXXVI
Rencontres de Moriond, March 17-24, 2001.

\bibitem{BBL}  G. Buchalla, A.J. Buras and M.E. Lautenbacher, Rev. Mod.
Phys. {\bf 68}, 1230(1996).

\bibitem{Ball}  P. Ball, JHEP {\bf 01}, 010 (1999).



\bibitem{GHLR}  M. Gronau, O.F. Hern\'{a}ndez, D. London and J.L. Rosner,
Phys. Rev. D{\bf 50}, 4529 (1994); {\it ibid.} {\bf 52}, 6356 (1995).



\bibitem{pipiCLEO}  CLEO Collaboration, D. Cronin-Hennessy, Phys. Rev. Lett.
{\bf 85}, 515 (2000).

\bibitem{CDF1}  CDF Collaboration, M. Tanaka, 7th International Conference
on B-Physics at Hadron Machines, Sea of Galilee, Kibbutz Maagan, Israel,
September 13-18, 2000.
\end{thebibliography}
\end{document}